\documentstyle[preprint,prl,aps,epsf,psfig]{revtex}
\tightenlines
\newcommand{\kmu}     {\mbox{$K^+ \! \rightarrow \! \mu^+ \nu_\mu$}}
\newcommand{\kmug}    {\mbox{$K^+ \! \rightarrow \! \mu^+ \nu_\mu \gamma$}}
\newcommand{\pnn}     {\mbox{$K^+ \! \rightarrow \! \pi^+ \nu \overline{\nu}$}}
\newcommand{\muplus}     {\mbox{$\mu^+$}}
\newcommand{\kplus}   {\mbox{$K^+$}}
\newcommand{\pizero}   {\mbox{$\pi^0$}}
\newcommand{\kmuthree}    {\mbox{$K_{\mu 3}$}}
\newcommand{\kmut}    {\mbox{$K^+ \! \rightarrow \! \pi^0 \mu^+ \nu_\mu$}}
\newcommand{\kpi}     {\mbox{$K^+ \! \rightarrow \! \pi^+ \pi^0$}}
\newcommand{\kpitwo}    {\mbox{$K_{\pi 2}$}}

\newcommand{\sdp}    {\mbox{${\rm SD}^+$}}
\newcommand{\sdm}    {\mbox{${\rm SD}^-$}}
\newcommand{\intp}    {\mbox{${\rm INT}^+$}}
\newcommand{\intm}    {\mbox{${\rm INT}^-$}}

\newcommand{\kmutwo}    {\mbox{$K_{\mu 2}$}}
\newcommand{\kmugu}   {\mbox{$K_{\mu\nu\gamma}$}}
\newcommand{\kegu}   {\mbox{$K_{e\nu\gamma}$}}
\newcommand{\chisq}    {\mbox{$\chi^2$}}

\begin{document}

%%%%%%%%%%%%%%%%%%%%%%%%%%%%%%%%begin text%%%%%%%%%%%%%%%%%%%%%%%%%%%%%%%%%%%%
\preprint{\vbox{\hbox{BNL-67217}
\hbox{PRINCETON/HEP/2000-2}
\hbox{TRI--PP--00--05}
\hbox{KEK Preprint 99-183}}}

\title{Measurement of Structure Dependent \kmug\ Decay}

%%%%%%%%%%%%%%%%%%%%%%begin author list%%%%%%%%%%%%%%%%%%%%%%%%%%%

\author{S.~Adler$^1$, M.S.~Atiya$^1$, I-H.~Chiang$^1$, M.V.~Diwan$^1$,
  J.S.~Frank$^1$, J.S.~Haggerty$^1$, S.H.~Kettell$^1$, 
  T.F.~Kycia$^1$\cite{AKS}, K.K.~Li$^1$, L.S.~Littenberg$^1$, 
  A.~Sambamurti$^1$\cite{AKS},  A.~Stevens$^1$,
  R.C.~Strand$^1$, C.~Witzig$^1$, 
T.K.~Komatsubara$^2$, M.~Kuriki$^2$,
  N.~Muramatsu$^2$, 
S.~Sugimoto$^2$, T.~Inagaki$^3$, S.~Kabe$^3$,
  M.~Kobayashi$^3$, Y.~ Kuno$^3$, T.~Sato$^3$, T.~Shinkawa$^3$,
  Y.~Yoshimura$^3$, Y.~Kishi$^4$, T.~Nakano$^4$, 
  M.~Ardebili$^5$, M.R.~Convery$^5$,
  M.M.~Ito$^5$, D.R.~Marlow$^5$,
  R.A.~McPherson$^5$, P.D.~Meyers$^5$, F.C.~Shoemaker$^5$,
  A.J.S.~Smith$^5$, J.R.~Stone$^5$, M.~Aoki$^6$\cite{masa},
  E.W.~Blackmore$^6$, P.C.~Bergbusch$^{6,8}$, D.A.~Bryman$^{6,8}$, 
A.~Konaka$^6$,
  J.A.~Macdonald$^6$, J.~Mildenberger$^6$, T.~Numao$^6$,
   P.~Padley$^6$, J.-M.~Poutissou$^6$,
R.~Poutissou$^6$,
  G.~Redlinger$^6$\cite{george},
P.~Kitching$^7$, and R.~Soluk$^7$ \\ (E787 Collaboration) }

\address{ 
$^1$Brookhaven National Laboratory, Upton, New York 11973\\
$^2$High Energy Accelerator Research Organization (KEK),
Tanashi-branch, Midoricho, Tanashi, Tokyo 188-8501, Japan \\ 
$^3$High Energy Accelerator Research Organization (KEK), 
Oho, Tsukuba, Ibaraki 305-0801, Japan \\ 
$^4$RCNP, Osaka University, 10--1 Mihogasaki, Ibaraki, Osaka 567-0047, 
Japan \\ 
$^5$Joseph Henry Laboratories, Princeton University, Princeton, 
New Jersey 08544 \\
$^6$ TRIUMF, 4004 Wesbrook Mall, Vancouver, British Columbia,
Canada, V6T 2A3\\
$^7$ Centre for Subatomic Research, University of Alberta, Edmonton,
Alberta, Canada, T6G 2N5\\
$^8$ Department of Physics and Astronomy, University of British Columbia, 
Vancouver, BC, Canada, V6T 1Z1
}

%%%%%%%%%%%%%%%%%%%%%%end author list%%%%%%%%%%%%%%%%%%%%%%%%%%%
\date{March 15, 2000}

\maketitle

\begin{abstract}
We report the first measurement of a structure dependent
component in the decay \kmug. 
Using the kinematic region where the muon kinetic energy is greater
than 137 MeV and the photon energy is greater than 90 MeV,
we find that the absolute value of the sum of the vector 
and axial-vector form factors is
$|F_V+F_A| =0.165 \pm 0.007 \pm 0.011$.
This corresponds to a branching ratio of
$BR(\sdp) = (1.33 \pm 0.12 \pm 0.18) \times 10^{-5}$.
We also set the limit $-0.04 <  F_V-F_A < 0.24$ at $90\%$ c.l.

\end{abstract}


\draft

The decay \kmug\ (\kmugu ) can proceed via two distinct 
mechanisms.
The first, internal bremsstrahlung (IB),
is a radiative version of the familiar \kmu\ (\kmutwo) decay:
its Feynman diagram has a photon emitted from the external
kaon or muon line.
The second, structure dependent radiative decay (SD),
involves the emission of a photon from intermediate states.
SD is sensitive to the electroweak structure of
the kaon and has been the subject of an extensive theoretical
literature\cite{old_theory,formalism}.  
In recent years most of this has been in the 
framework
of Chiral Perturbation Theory (ChPT)\cite{bijnens}.
The differential rate in the \kplus\ rest frame
can be written \cite{formalism} in terms of 
$x\equiv{2 E_\gamma \over M_K}$ and 
$y\equiv{2 (E_\mu + M_\mu) \over M_K}$, 
where $E_\gamma$ is the photon energy, $E_\mu$ is the muon kinetic energy, 
$M_\mu$ is the \muplus\ mass, and $M_K$ is the \kplus\ mass:
\begin{eqnarray*}
%{d \Gamma_{\kmugu} \over dxdy}& =&  A_{IB} f_{IB} (x,y) \\
{d \Gamma_{K_{\mu\nu\gamma}} \over dxdy}& =&  A_{IB} f_{IB} (x,y) \\
& + & A_{SD}\big[(F_V+F_A)^2 f_{SD^+}(x,y) +
             (F_V-F_A)^2 f_{SD^-}(x,y) \big] \\
& - &  A_{INT}\big[(F_V+F_A) f_{INT^+}(x,y) + 
             (F_V-F_A) f_{INT^-}(x,y) \big],
\end{eqnarray*}
where
\begin{eqnarray*}
f_{IB}(x,y)& = & \bigg[ {1-y+r \over x^2(x+y-1-r)}\bigg] \\
           &   & \times \bigg[x^2+2(1-x)(1-r) -
               {2xr(1-r) \over x+y-1-r}\bigg],\\
f_{SD^+} & = & \big[x+y-1-r \big]\big[(x+y-1)(1-x)-r\big], \\
f_{SD^-} & = & \big[1-y+r \big]\big[(1-x)(1-y)+r\big], \\
f_{INT^+}& = & \bigg[ {1-y+r \over x(x+y-1-r)} \bigg]
               \big[(1-x)(1-x-y)+r \big],\\
f_{INT^-}& = & \bigg[ {1-y+r \over x(x+y-1-r)} \bigg]
               \big[x^2-(1-x)(1-x-y)-r\big],
\end{eqnarray*}

\begin{eqnarray*}
%x & = & {2 E_\gamma \over M_K},\\
%y & = & {2 E_\mu \over M_K},\\
r & = & \bigg[{M_\mu \over M_K }\bigg]^2, \\
A_{IB} & = & \Gamma_{K_{\mu2}} {\alpha \over 2 \pi} {1 \over (1-r)^2},\\
A_{SD} & = & \Gamma_{K_{\mu2}} {\alpha \over 8 \pi} {1 \over r(1-r)^2}
                  \bigg[ {M_K \over F_K} \bigg]^2,\\
A_{INT} & = & \Gamma_{K_{\mu2}} {\alpha \over 2 \pi} {1 \over (1-r)^2} 
                  {M_K \over F_K}.\\
\end{eqnarray*}
In these formulas, 
$F_V$ is the vector form factor,
$F_A$ is the axial form factor \cite{ffs},
$\alpha$ is the fine structure constant (1/137.036),
$F_K$ is the \kplus\ decay constant ($159.8 \pm 1.4 \pm 0.4$ MeV), and
$\Gamma_{K_{\mu2}}$ is the width of the \kmutwo\ decay.

\begin{figure}
\centerline{\psfig{figure=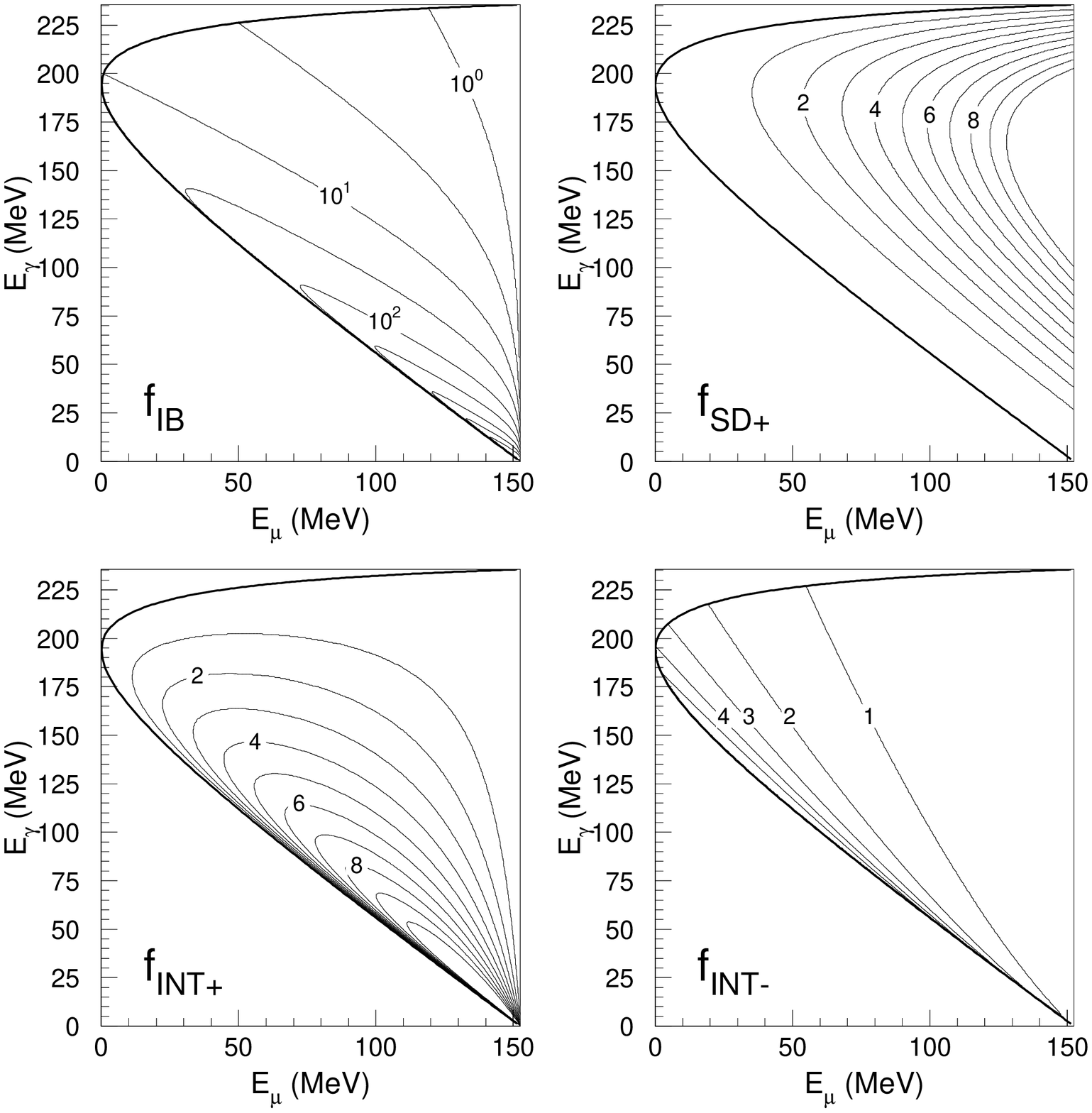,height=3in}}

\caption{\label{fig:plot_wmug} Spectral shape functions
for IB, \sdp, \intp and \intm\ components of \kmugu.
The normalizations are arbitrary, and the scale on the $f_{IB}$ 
plot is logarithmic. 
The \sdm\ component is not shown because 
it peaks at low muon momentum and has negligible effect 
on the current analysis.}
\end{figure}

\sdp\ and \sdm\ refer to different photon
polarizations and these components do not mutually interfere.
Both \sdp\ and \sdm\ can interfere with IB, however, resulting
in the terms labeled \intp\ and \intm .
Figure \ref{fig:plot_wmug} shows the shapes of 
$f_{IB}$, $f_{SD^+}$, $f_{INT^+}$ and $f_{INT^-}$.
The \sdp\ component peaks at high muon and photon energy, making it the easiest 
of the SD components to observe.
This analysis, therefore, is mostly aimed at observing the \sdp\ component.
The form factors of the decay, $F_V$ and $F_A$, can,
in principle, depend on $q^2$, which is given by $q^2=M_K^2 - 2M_KE_{\gamma}$
in the \kplus\ rest frame.
In an ${\cal O}(p^4)$ ChPT calculation
\cite{bijnens,explain}, however, they are found to be $q^2$ independent and 
are given by $F_V + F_A = 0.137$, $F_V - F_A = 0.052$, which
corresponds to $BR(\sdp) = 9.22 \times 10^{-6}$.
In the data analysis, we initially assume that they are constant, then 
test for $q^2$ dependence.

% Fig 1 was here

The IB component of \kmugu\ has been well measured in other experiments 
and found to agree with the QED prediction \cite{akiba}.
The structure dependent components, on the other hand, 
have not yet been measured.
For the \sdp\ component, the best limit is 
BR(\sdp) $< 3.0 \times 10^{-5}$ \cite{akiba}.
There is also a limit on the combination BR(\sdm + \intm ) 
$< 1.3 \times 10^{-3}$ \cite{akiba}.  In terms of the form factors, 
these limits translate into
$|F_V + F_A| < 0.23$, $-0.3 < (F_V - F_A ) < 2.5$.  

Better measurements
are available from the closely related process $K^+ \to e^+ \nu \gamma$
(\kegu).
In \kegu, the IB term is heavily suppressed by helicity, 
so that the SD terms are easier to extract.
Since the INT terms are also highly suppressed, the signs of
the form factors are practically impossible to measure.
In ${\cal O}(p^4)$ ChPT, $F_V$ and $F_A$ for \kegu\ are identical to those
for \kmugu.
The \kegu\ experiments\cite{keng} give $|F_V + F_A| = 
0.148 \pm 0.010$, $|F_V - F_A| < 0.49$, in agreement with ${\cal O}(p^4)$ ChPT.

In \kmugu, the IB term is large, thus complicating the extraction of 
the SD terms, but also making the INT terms comparable in size 
to the SD terms.
This makes it possible, in principle, to measure the sign as well as the
magnitude of the form factors.
In addition to its potential for checking the predictions of ChPT,
\kmugu\ is also interesting as a probe of non-Standard Model 
CP-violation\cite{Tviol}.  
One can look for a T-violating component of
muon polarization transverse to the plane of the decay.
Such an effect is proportional to the INT components.

\begin{figure}
\centerline{
\hfill
\psfig{file=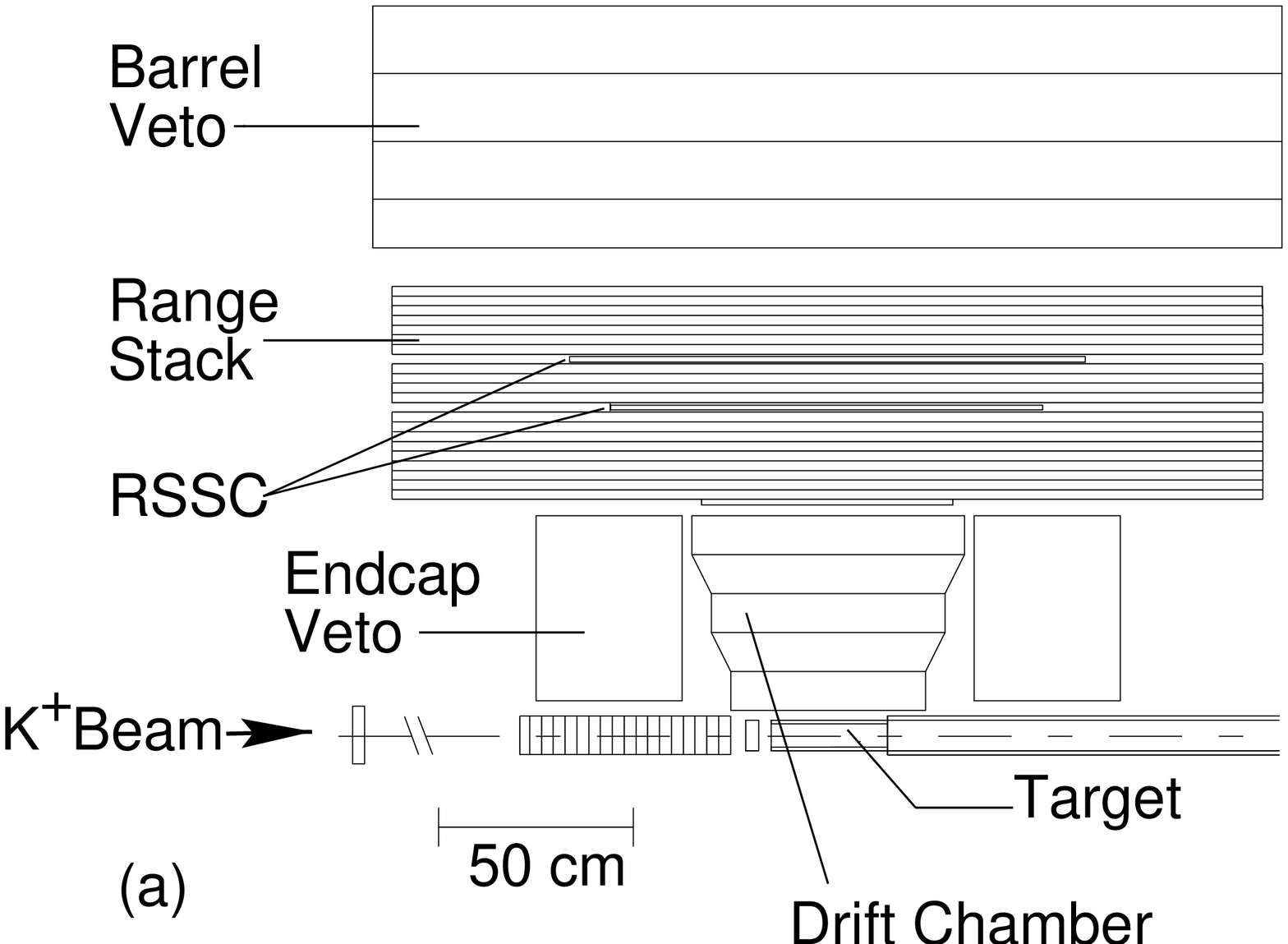,height=1.5in}
\hfill
\psfig{file=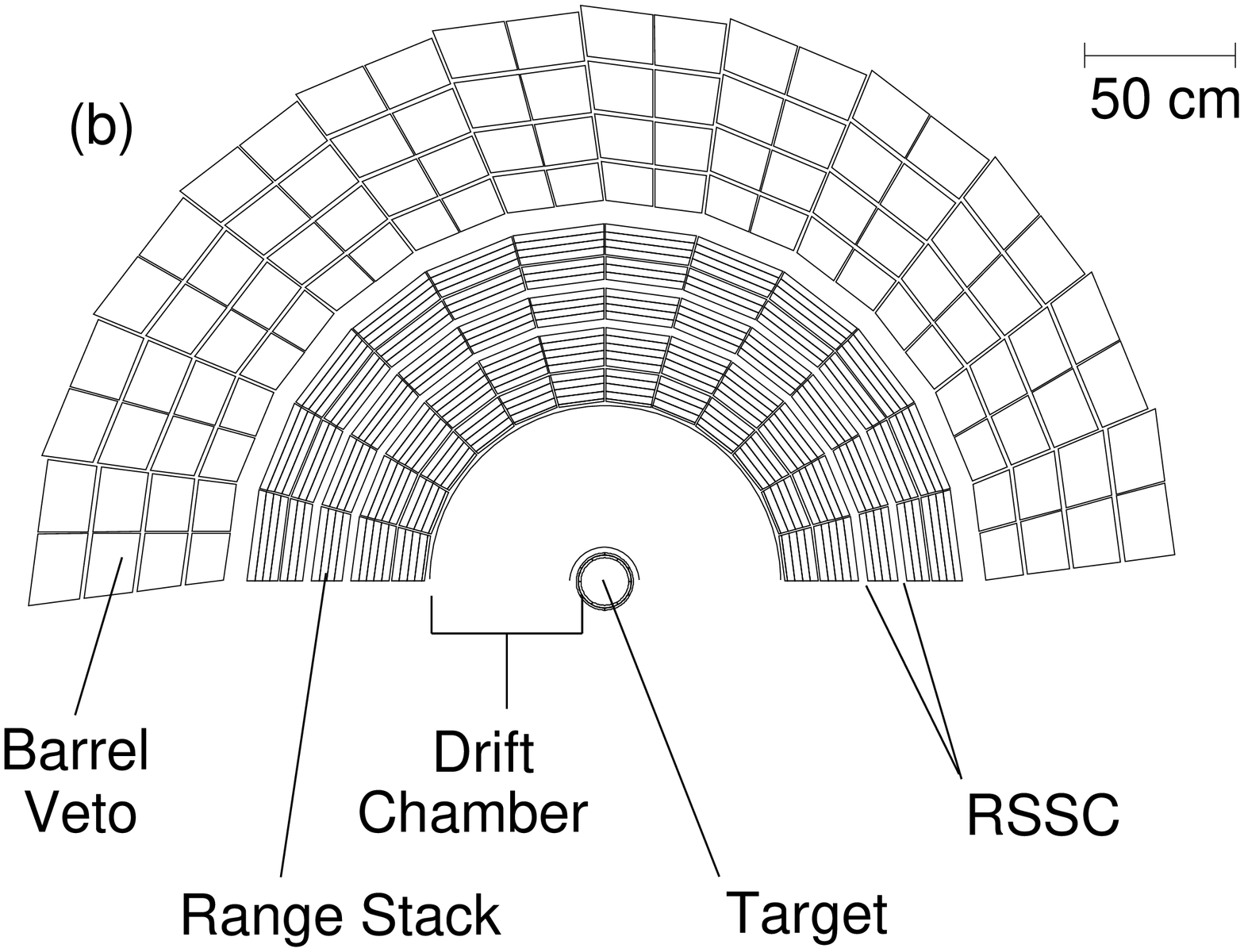,height=1.7in}
\hfill
}
\caption{\label{fig:e787}
Side-view (a) and end-view (b) of 
upper half of the E787 detector.}
\end{figure}

The E787 experiment at the Brookhaven Alternating Gradient Synchrotron (AGS)
\cite{nim},
shown schematically in Figure \ref{fig:e787}, 
was used to look for the \sdp\ component.
E787, originally designed to search for \pnn, uses a beam of
\kplus\ mesons brought to a stop in a scintillating fiber target.
From there, charged decay products can enter the drift chamber
where their momentum is measured in a 1-T magnetic field.
The charged tracks then enter the Range Stack (RS), 
which consists of 21 layers of
scintillator and two layers of straw tube chambers (RSSC).
Most tracks range out in the RS, 
thus allowing measurements of their total energy
and range.
A 4$\pi$ photon detection system, composed of the Barrel Veto (BV) 
and two endcaps, surrounds the central region.
In the present application the BV, covering
70\% of the solid angle and composed of lead and scintillator, 
is used to detect the photons of interest
as well as to rule out the presence of more than one photon.  

The \kmugu\ data was taken with the upgraded E787 detector, which was completed in
1994.
In this analysis, the redundant charged track energy 
and momentum measurements are
combined (assuming a \muplus\ mass) to give an improved measurement of
the track kinematics.
The rms resolution of this combined quantity is $\sigma_{p_\mu}=0.0164 \cdot p_{\mu} -
0.86\,$MeV/$c$, for $205 < p_\mu < 236$ MeV/$c$, 
where $p_\mu$ is the combined measurement expressed as a momentum.
The resolutions for the azimuthal ($\phi$) and polar 
($\theta$, with respect to the beam) angles of the
muons are each $32$ mrad.
The resolutions on the photon kinematic quantities are
$\sigma_{E_\gamma} = 1.676 \sqrt{E_\gamma}$ MeV 
($E_\gamma$ in MeV), $\sigma_{\phi} =
25\,$mrad, and $\sigma_{\theta} = 45\,$mrad.

% Fig 2 was here

A special trigger designed to search for the 
\sdp\ component of \kmugu\ required a high energy charged track in 
the central region, a high energy photon in the BV, and 
no other photons in the event. 
A two-day run using this trigger netted a total exposure
of $9.2 \times 10^9$ \kplus\ , yielding a total of 
$1.5 \times 10^6$ \kmugu\ triggers.

Analysis of the events passing the trigger proceeds in three 
steps: event reconstruction, background rejection, and \kmugu\ spectrum
fitting.
In the reconstruction step, the energy, time, and flight direction
of the charged track and photon are calculated. 
Any additional photon energy not associated with the primary photon
is also recorded. 
A kinematic fit to the \kmugu\ hypothesis is applied to the charged 
track and the photon.
Since there are four constraints (conservation of momentum and energy) and
three unmeasured quantities (momentum of the neutrino), the kinematics
are over-constrained and non-\kmugu\ events should have a bad fit \chisq.
Additionally, the kinematic fit yields measurements of $E_{\mu}$ and 
$E_{\gamma}$ with better resolution than the raw quantities.
These are the variables that are used in the final spectrum fits.

The two main types of background that need to be rejected are
\kmutwo\ accompanied by an accidental photon and \kmut\ (\kmuthree)
or \kpi\ (\kpitwo) where one of the photons from the \pizero\ 
decay satisfies the photon requirement and the other photon is undetected. 
The \kmutwo+accidental background can be suppressed in two independent 
ways: by requiring a tight time coincidence between the muon and the
photon and by examining the kinematics of the decay.
Since the accidental photon is randomly oriented relative to the 
muon, the cut on the \chisq\ of the kinematic fit to the 
\kmugu\ hypothesis is especially effective against this background.
Both \kmuthree\ and \kpitwo\ background can also be rejected in two 
independent ways: vetoing on any additional photon energy in the event
and by the kinematics of the decay.
The requirement that the charged track energy be above the \kmuthree\ endpoint
($E_{\mu^+} > 137$ MeV) is especially effective against this type of background.

Since both types of background can be rejected by two independent methods,
the total rejection of all cuts for each background
may be calculated as the product of the rejections of the two methods. 
This allows a calculation of expected background based solely on the
data, thus lowering the estimated systematic error.
In the final signal region defined by $E_{\mu^+} > 137$ MeV and 
$E_{\gamma} > 90$ MeV, the expected background (with statistical error) 
from the \kmutwo+accidental source is 79.4 $\pm$ 4.8 events.
The \kmuthree\ and \kpitwo\ backgrounds are treated together and give a
total expected background of 25.2 $\pm$ 3.8 events.

\begin{figure}
\centerline{\psfig{figure=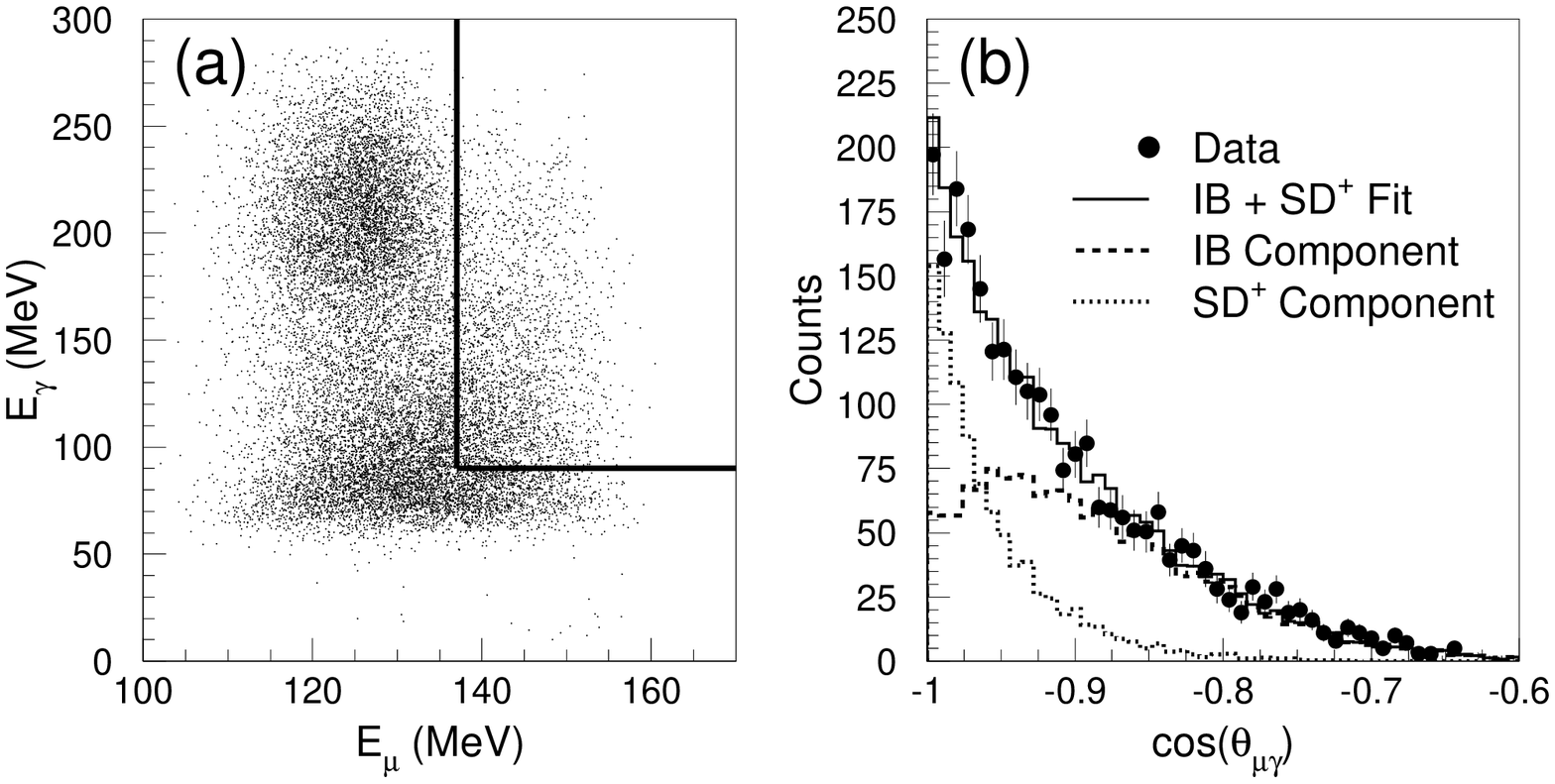,height=1.5in}}
\caption{\label{fig:sprocl_anal}
(a) Spectrum of events passing all but the final kinematic cut. 
At this point, nearly all events are \kmugu\ except those in the 
region  $E_\mu < 137$ MeV and $E_\gamma > 150$ MeV, where background 
from \kmuthree\ and \kpitwo\ is concentrated.
The IB component of \kmugu\ is concentrated at $E_\gamma < 100 $ MeV. 
The box marks the final cut of $E_\mu > 137$ MeV and $E_\gamma > 90$ MeV,
within which the \sdp\ component is enhanced.
(b) Counts vs. $\cos(\theta_{\mu\gamma})$ and various fits as described in text.}
\end{figure}

Figure \ref{fig:sprocl_anal}(a) shows the final spectrum of events with
the final signal region in the upper right corner delineated by the solid line.
The number of events in this  region is 2693, the vast majority of which are
\kmugu.
As a simple way of testing whether the \kmugu\ events are consistent with
being only IB, we examine the distribution of the opening angle 
between the muon and the photon ($\cos\theta_{\mu\gamma}$).
Figure \ref{fig:sprocl_anal}(b) shows this distribution for 
background-subtracted data.
Superimposed on the data are Monte Carlo distributions for IB and 
\sdp  components of \kmugu.
When only an IB component is allowed, the quality of the fit is very poor 
(\chisq = 300, with 48 degrees of freedom).
When an \sdp\ contribution is allowed, a much better fit is obtained
(\chisq = 58)
\cite{cosop_fit},
clearly indicating that a structure dependent component is present.

\begin{figure}
\centerline{\psfig{figure=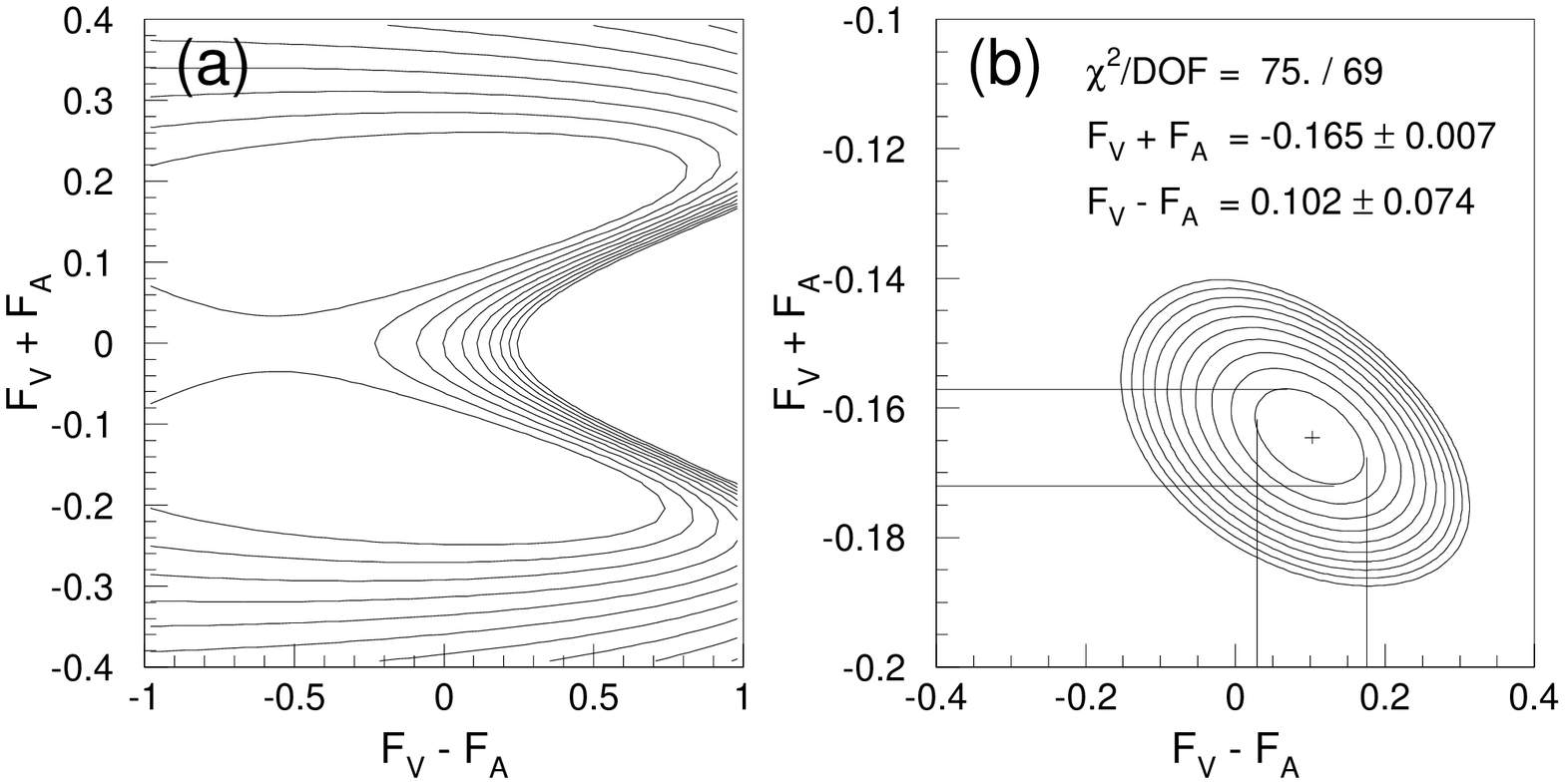,height=2.0in}}
\caption{
\label{fig:fvfa}
\chisq\ contours for the fit
to the $E_{\mu^+}$ vs. $E_{\gamma}$ distribution.
(a) Contours for all plausible
values of the form factors. Each contour represents 50 units of \chisq .
(b) Near a \chisq\ minimum. 
In this plot, each contour corresponds to one unit of \chisq .  
The one-standard-deviation uncertainties
for $F_V +F_A$ and $F_V -F_A$ 
are also shown.}
\end{figure}

The fit is incomplete, however, because it does not include the 
effects of the other \kmug\ components (\sdm, \intp, \intm).
To include these effects, we generate
Monte Carlo distributions with SD and INT components weighted by the form
factors and normalized to the IB component. 
In Figure \ref{fig:fvfa}, we plot the \chisq\ between
the $E_{\mu^+}$ vs. $E_{\gamma}$ histogram of this Monte Carlo sample and
that observed in data (after background subtraction) as a function of the form factors.
The minimum \chisq\ is 75 with 69 degrees of freedom. 
%we plot the \chisq\ between
%data and Monte Carlo of the
%$E_{\mu^+}$ vs. $E_{\gamma}$ distribution 
%as a function of the form factors.
%The form factors are normalized to the IB spectrum described above.
%As shown in Figure \ref{fig:fvfa_laur}, 
The best fit values are
$$
 |F_V+F_A| = 0.165 \pm 0.007,\quad  F_V-F_A = 0.102 \pm 0.073,
$$
where the errors are statistical.
The minimum \chisq's found in the regions where $F_V+F_A < 0$ and where
$F_V+F_A > 0$ differ by only 0.2.
We thus have no information about the sign of $F_V+F_A$ and can
only measure its absolute value.
The result corresponds to a branching ratio 
of BR(\sdp) $= (1.33 \pm 0.12) \times 10^{-5}$.

The largest systematic errors associated with the form factor measurements come
from possible distortions of the \kmugu\ spectrum induced by differences between
the true detector and the Monte Carlo simulation.
The two largest sources of distortion are 
non-linearity in the measurement of the photon energy and 
uncertainty in the thickness of the individual RS scintillator layers.
For  $|F_V+F_A|$, these two sources lead to uncertainties of 0.0095 and 0.0054, 
respectively. 
For  $|F_V-F_A|$, they are 0.028 and 0.033.
The systematic errors due to uncertainty in the level of background present
in the final sample are estimated in the data-based background studies 
described above.
They are found to be very small, totalling 0.0007 for $|F_V+F_A|$ and 0.0097 for
$F_V-F_A$.
Even a much enhanced background level would have only a small effect on the 
measurements. 
Adding the individual errors in quadrature, we find a total 
systematic error of 0.011 for 
$|F_V+F_A|$ and 0.044 for $F_V-F_A$

As a check on possible systematic errors, the branching ratio for
the IB component has also been extracted.  
This was accomplished by normalizing to a sample of \kmutwo\ decays that 
was taken simultaneously with the \kmugu\ data.
For $E_{\mu} > 100\,$MeV and
$E_{\gamma} > 20\,$MeV, we find BR(IB)$ = (3.6 \pm 0.3) \times
10^{-4}$, in good agreement with the theoretical value for this
kinematic region, $3.3 \times 10^{-4}$.
Other checks included changing the binning of the 
$E_{\mu^+}$ vs. $E_{\gamma}$ histogram and varying the $E_\gamma$ cut. 
While both of these checks were limited by statistics, neither showed a 
systematic trend as the parameters were varied.
Therefore, no systematic error is associated with these effects.

As mentioned above, the form factors $F_V$ and $F_A$ have, to this point, 
been considered independent of $q^2$ .
To assess the effect of including $q^2$ dependence, we assume the 
following form factor form:
$$
F_V = {F_V(q^2=0) \over 1-q^2/m_V^2},\quad
F_A = {F_A(q^2=0) \over 1-q^2/m_A^2}.
$$ 

We take $m_V=0.870$ GeV (the $K^*$ mass) and $m_A=1.270$ GeV (the $K_1$ mass)
and refit the measured \kmugu\ spectrum in terms of the parameters
$F_V(0)+F_A(0)$ and $F_V(0)-F_A(0)$.
The best fit parameters are:
$$
|F_V(0) + F_A(0)| = 0.155  \pm 0.008,\quad
F_V(0) - F_A(0) = 0.062  \pm 0.078.
$$
The corresponding \sdp\ branching ratio is $(1.37 \pm 0.12) \times 10^{-5}$.
Although the value of $|F_V(0) + F_A(0)|$ differs somewhat from that
obtained assuming $q^2$ independence, the associated branching ratio 
changes only slightly.
Furthermore, the minimum \chisq\ of the fit is very insensitive to 
$m_V$ and $m_A$, so we are unable to measure them and cannot offer 
evidence of $q^2$ dependence.

% Fig. 4 was here

In conclusion, we have observed a structure dependent
component in the decay \kmug. 
Under the assumption of $q^2$ independence, the associated 
form factors are
$$
|F_V + F_A| = 0.165 \pm 0.007 \pm 0.011,\quad
 F_V - F_A = 0.102 \pm 0.073 \pm 0.044.
$$
Since the measurement of $F_V-F_A$ is not significantly different from
zero, we add statistical and systematic errors in quadrature and 
calculate the 90\% confidence level:
$$
 -0.04 < F_V - F_A < 0.24.
$$
The $|F_V+F_A|$ measurement is consistent with the
previous result on $K^+ \to e^+ \nu \gamma$, but disagrees with
the ${\cal O}(p^4)$ ChPT prediction by about two standard deviations.
This is perhaps not surprising since at 
higher order in ChPT, kaon form factors are expected to differ from 
those of the pion
\cite{Ametller}. 
The  ${\cal O}(p^6)$ calculation has been done for pions 
\cite{Bijnens:1997wm}, but not yet for kaons.
The limit on $F_V - F_A$ is consistent with  ${\cal O}(p^4)$ ChPT
and is  significantly better than any previously obtained from kaon decay. 
A more detailed description of the analysis can be found
in reference \cite{convery}. 

We gratefully acknowledge the dedicated effort of the technical
staff supporting this experiment and of the Brookhaven AGS
Department.  This research was supported in part by the
U.S. Department of Energy under Contracts No. DE-AC02-98CH10886,
W-7405-ENG-36, and grant DE-FG02-91ER40671, by the Ministry of
Education, Science, Sports and Culture of Japan
through the Japan-U.S. Cooperative Research Program
in High Energy Physics and under the Grant-in-Aids for
Scientific Research, for Encouragement of Young Scientists and for
JSPS Fellows,
and by the Natural
Sciences and Engineering Research Council and the National Research
Council of Canada.

%%%%%%%%%%%%%%%%%%%%%%%%%%%%%%begin references%%%%%%%%%%%%%%%%%%%%%%%%%%%%%%%%%

%%%%%%%%%%%%%%%%%%%%%%%%%%%%%%end references%%%%%%%%%%%%%%%%%%%%%%%%%%%%%%%%%

%\vskip 1in

%\vskip 1in

%\vskip 1in

\begin{references}
\bibitem{ffs}We use the convention of the Particle Data Group\cite{PDG},
where $F_V$ and $F_A$ are dimensionless and larger by a 
factor of $\sqrt{2}$ than in many theoretical papers and where a minus sign
precedes the interference term, thus changing their sign.
\bibitem{explain}At ${\cal O}(p^4)$ in ChPT, these form factors are closely
related to those of $\pi^+ \! \rightarrow \! \ell^+\nu_{\ell}\gamma$. 
Thus, the vector form factor is predicted via CVC from the $\pi^0$ lifetime to 
be $F_V=0.095 \pm 0.002$ and the axial form factor from the $F_A$ measured in 
$\pi^+ \! \rightarrow \! e^+\nu_e\gamma$, $F_A=0.043 \pm 0.006$.
\bibitem{cosop_fit} The best-fit \sdp\ branching ratio in the 
$\cos(\theta_{\mu\gamma})$ fit is
$BR(\sdp) = (1.37 \pm 0.12)\times 10^{-5}$.
\end{references}
\end{document}